# Magnetic Pitch as a Tuning Fork for Superconductivity

Yishu Wang[1,2], Yejun Feng[1,3], J.-G. Cheng[4], W. Wu[4], J. L. Luo[4], T. F. Rosenbaum[1,2]

[1]The James Franck Institute and Department of Physics, The University of Chicago, Chicago, Illinois 60637, USA
[2]Division of Physics, Mathematics, and Astronomy, California Institute of Technology, Pasadena, California 91125, USA
[3]The Advanced Photon Source, Argonne National Laboratory, Argonne, Illinois 60439, USA
[4]Beijing National Laboratory for Condensed Matter Physics and Institute of Physics, Chinese Academy of Sciences, Beijing 100190, P. R. China

**From the lodestone-based compass to modern theories of phase transitions [1], magnetic materials have played an outsized role in revealing the shape of the world around us. The similarly venerable field of superconductivity serves as a prime example of emergent, collective behavior in nature, with raised hopes of technological import with the discovery of exotic superconducting order in the cuprates. Magnetism and superconductivity often compete for preeminence as a material's ground state, and in the right circumstances the fluctuating remains of magnetic order can induce superconducting pairing. The intertwining of the two on the microscopic level, independent of lattice excitations, is especially pronounced in heavy fermion compounds, rare earth cuprates, and iron pnictides. Here we point out that for a helical arrangement of localized spins, a variable magnetic pitch length provides a unique tuning process from ferromagnetic to antiferromagnetic ground state in the long and short wavelength limits, respectively. Such chemical or pressure adjustable helical order naturally provides the possibility for continuous tuning between ferromagnetically and antiferromagnetically mediated superconductivity. At the same time, phonon mediated superconductivity is suppressed because of the local ferromagnetic spin configuration [2]. We employ synchrotron-based magnetic x-ray diffraction techniques [3-5] to test these ideas in the recently discovered superconductor, MnP [6]. This sensitive probe directly reveals a reduced-moment, helical spin order at high pressure proximate to the superconducting state, with a tightened pitch in comparison to that at ambient pressure where superconductivity is absent. The correlation between magnetic pitch length and superconducting transition temperature in the (Cr/Mn/Fe)(P/As) family [7-13] suggests a strategy for using spiral magnets as interlocutors for spin fluctuation mediated superconductivity.**

High-pressure electrical studies of the transition metal compound MnP [6] manifest a complex pressure-temperature (*P*-*T*) phase diagram (Fig. 1). At ambient pressure, MnP develops helical spin order below $T$ = 50K with a wave vector $\boldsymbol{Q}$=(0.117, 0, 0) [7]. Under pressure, the helical order is quickly replaced by ferromagnetism at approximately 1GPa, and another magnetic state, assumed to be antiferromagnetic [6], emerges for $P > 2$GPa. Superconductivity appears after the high-pressure magnetic phase is suppressed at $P \sim 7$GPa [6]. No details of the spin structure in the high-pressure magnetic phase have been obtained.

To elucidate the cascade of magnetic states in the *P-T* phase diagram, and their relation to superconductivity, we performed non-resonant x-ray magnetic diffraction under pressure [3-5], and discover helical magnetic order with ***Q'***~(0.25, 0, 0) presaging the high-pressure superconductor (Fig. 2). We observe a pair of superlattice peaks in mirror symmetry to the lattice order at three pressures, 3.17, 5.28, and 6.43 GPa, but absent at $P$ = 8.99 and 10.4 GPa (Extended Data Fig. 1). These diffraction peaks are always of single crystal nature (Extended Data Fig. 2) and their pressure evolution manifests a compressibility commensurate to that of the *a*-axis. The low transferred momentum of (1-*Q*', 0, 0) rules out diffraction from integer lattice orders from both MnP and other components of the high-pressure cell (diamond and Ag manometer) [4-5]. The peak intensities lie in the range of 1-4 $10^{-8}$ relative to the (200) lattice intensity, which are too weak for typical charge order diffraction, but comparable with diffraction intensities of the low-pressure helical order observed under the same condition (Extended Data Fig. 1). Our diffraction results are insufficient to fully refine the high-pressure spin structure. However, it is reasonable to identify it as helical order with tightened pitch (H*a*-II, Fig. 1 inset), thereby providing a consistent perspective on all three spin structures (H*a*-I, FM, and H*a*-II). The spiral magnetism develops with a varying twist angle between neighboring spin pairs along the wave vector direction, a subtle result due to pressure-dependent, competing exchange constants from multiple close neighbors in an anisotropic lattice [14].

The boundary of the magnetic phase is determined most accurately by the pressure evolution of the lattice. Single crystal refinement of five to six Bragg orders of MnP at each pressure indicates that the lattice structure remains in the orthorhombic phase to 10.4 GPa. Longitudinal scans of lattice orders such as (200), (021), (220), and (222), showing instrument resolution limited profiles with no noticeable peak splitting, support this conclusion. All three lattice constants evolve nonlinearly at low pressure but linearly at high pressure, with the crossover defining the critical pressure, $P_c$=6.7 ± 0.2 GPa (Fig. 3), consistent with the range where magnetic diffraction was observed directly. The lattice changes continuously under pressure to a sensitivity level of $|\Delta l|/l \sim 1\ 10^{-3}$ (Fig. 3). The orthorhombic structure of MnP is considered to be a distortion from the hexagonal structure of NiAs [15], as the two symmetries can evolve continuously across the ratio $a/c$=1.732. Under pressure, the orthorhombic distortion in MnP, measured by $a/c$, keeps increasing and moves away from the hexagonal symmetry (Extended Data Fig. 3). While helical order in both MnSi and CrAs are suppressed by pressure through a clear first-order quantum phase transition [11, 12, 16], the quantum phase transition in MnP at $P_c$ is isostructural and could be continuous.

The lattice evolution with pressure indicates a significant magnetostriction, which can be extracted from $\Delta c$ and $\Delta a$ of the lattice and scaled to the magnetic phase boundary $T_{C/N}$ as $\Delta c \sim \Delta a \sim T_{C/N}$ (Fig. 3c), regardless of whether there is underlying ferromagnetic or antiferromagnetic order. Since the magnetostriction $\Delta l$ is directly related to the staggered magnetic moment $\langle m \rangle$, it goes to zero at $T_{C/N}$. Beyond $P_c$, an energy density of 7 GPa distributed over eight valence electrons in the P 3*p* and Mn 3*d* orbitals [17] increases the electron kinetic energy *t* by ~15 meV/electron, comparable to the magnetic exchange constants *J* (2.5-11 meV [18]). An increasing $t/J$ ratio reduces the ordered moment $\langle m \rangle$ and eventually destabilizes the magnetism. While $\langle m \rangle$ drops to zero at a quantum critical point, the fate of individual local moments remains of high interest, as exemplified in heavy fermion materials [19].

Spins in MnP are deep in the local limit at ambient pressure given a Rhodes-Wohlfarth ratio of 2.2 (Extended Data Fig. 4). The 15 meV/electron increase in kinetic energy sufficient to destabilize the magnetic order is not enough to fully delocalize the 3*d* moments, considering their 0.20 eV bandwidth [17]. Therefore, MnP is a system with local moments surviving beyond the quantum critical point, and spin fluctuations in the disordered state naturally raise special interest about magnetically driven superconductivity.

In the disordered phase, the predominant spin fluctuation modes likely are still dictated by the nearby magnetic instability [19-21]. In MnSi, helical fluctuations in the form of spiral/helix paramagnons were observed for $T > T_c$ despite a weak first order transition. Those fluctuations center at a wave vector similar in magnitude to the ordering wave vector *Q*, but with a random direction [21], presumably because of the short range Dzyaloshinskii-Moriya interaction in a cubic lattice symmetry. In MnP and CrAs, the lattice anisotropy likely confines wave vector directions of magnetic fluctuations. With no significant evolution of *Q* under pressure (Fig. 2) [13], their disordered phases should possess spin fluctuations dominated by the magnetic instability in the ordered phase, *i.e*. spiral modes centered at $Q \sim (0.25, 0, 0)$ and (0.36, 0, 0) for MnP and CrAs respectively.

These spiral modes are particularly interesting in terms of the competition between spin and lattice (phonon) fluctuations and their connection to superconducting pairing of *s*, *p*, or *d* character. Consider a helical fluctuation at a finite wavelength. By contrast to the usual antiferromagnet, spins of nearest neighbors along the wave vector $\mathbf{Q}$ direction share a large ferromagnetic projection. This suppresses phonon-mediated superconductivity due to on-site pairing of itinerant electrons [2], amplifying the effect of the spin fluctuations. Furthermore, varying the pitch of the helical order provides a continuous tuning of local ferromagnetic order versus intermediate-range antiferromagnetic order, thus tilting the competition between the two types of magnetically mediated superconductivities.

The scenario of ferromagnetic to antiferromagnetic superconductivity facilitated by different pitches of helical fluctuations can be evaluated directly in the series of MnSi, MnP, and CrAs, in light of their different magnetic wave vectors (Fig. 4a). With a small wave vector (a long pitch) of (0.017, 0.017, 0.017) [21, 22], MnSi does not superconduct under pressure down to at least 10 mK [16]. This is also true for MnP at low pressure, where the helical order with a wave vector of (0.117, 0, 0) [7] was simply replaced by a ferromagnetic order at $P$ ~ 1 GPa, and no superconductivity was observed down to 350 mK [6]. On the other hand, both MnP at high pressure (0.25 r.l.u.) and CrAs (0.36 r.l.u.) have relatively large wave vectors (short pitches) and demonstrate superconducting ground states once the helical order is suppressed by pressure (Fig. 4a) [11, 12].

Our focus on local moment helical order complements itinerant models of continuous tuning by band filling from ferromagnetic to antiferromagnetic order with a concomitant switch between magnetically mediated superconductivities of different symmetries [23]. Through the comparison of the cuprates and $Sr_2RuO_4$, it appears that ferromagnetically mediated superconductivity typically has an orders of magnitude lower transition temperature than its antiferromagnetic analogue of the same dimensionality [23]. For three-dimensional helical magnets such as MnP and CrAs with $T_c$ = 1 to 2 K (Fig. 4a), the corresponding ferromagnetic type could be below the lowest range of temperatures measured to date.

The issue of pairing symmetry is more tenuous, but the model of helical magnets allows certain predictions. Along the wave vector direction of helical spin fluctuations, two antiferromagnetically-coupled itinerant electrons are separated by a distance $r = \lambda/2$ of the helical order (Fig. 4b). The facts that $r$ is non-zero over the helical wave and that $Q$ mandates a preferred axial direction suggest that the superconductivity state might be of the singlet $d_{z^2}$ type, especially in light of the low symmetry lattice structures of MnP and CrAs. The distance $r$ is necessarily smaller than the mean free path of itinerant electrons, thereby allowing the electron pair overlap to maintain phase coherence without being particularly sensitive to disorder (Fig. 4b). In MnSi, with a long wavelength of 180 Å [22], itinerant electrons have to antiferromagnetically couple over a distance $r$ ~ 90 Å, so that the phase coherence between electrons could become tenuous. Instead, the local ferromagnetic fluctuations dominate, and the potential superconductivity could be of $p$-wave type at a much lower temperature.

Dimensionality of the spin fluctuations is another interesting issue. The helical order in 3$d$ compounds can be compared with incommensurate antiferromagnetic order in heavy fermion materials like $CeCu_{6-x}Au_x$ [19], where spin fluctuations with 2D character were observed around the ordering wave vector $\boldsymbol{Q}$ [20]. Even though the effective low-dimensionality enhances the spin fluctuations, the extremely low magnetic coupling strength in $CeCu_{6-x}Au_x$ [19] suppresses the possible magnetic mediated superconductivity below experimental sensitivity. Spin fluctuations in MnP are likely 3D judging from the $T^{3/2}$ dependence of the resistivity [6], but they are matched with a large magnetic coupling strength [18] and bandwidth [24], so $T_c$ could still be measurable even at a level of $T_N$/1000. The 3*d* helical magnets of the (Cr/Mn/Fe)(P/As) family [7-13] thus present manifest opportunities to further our understanding of the linkage between magnetism and unconventional superconductivity.

**Acknowledgments**
We thank P.B. Littlewood for helpful conversations. We are also grateful for assistance by B. Fisher on magnetization measurements, and for sample preparation at the MRSEC facilities of the University of Chicago supported by NSF. The work at the California Institute of Technology was supported by the NSF Division of Materials Research. The use of the Advanced Photon

Source and the Center for Nanoscale Materials of Argonne National Laboratory was supported by the U.S. Department of Energy Office of Science User Facilities. J.-G.C. and J.L.L. were supported by the MOST and NSF of China, and the Strategic Priority Research Program of the Chinese Academy of Sciences.

**Author contributions**

Y.F. and T.F.R. designed the research. J.-G.C., W.W. and J.L.L. provided samples. Y.W. and Y.F. performed measurements and developed the theoretical framework. Y.W., Y.F., and T.F.R. prepared the manuscript. All authors commented.

**Additional information**

Correspondence and requests for materials should be addressed to Y.F. (yejun@anl.gov) or T.F.R. (tfr@caltech.edu).

**Competing financial interests**

The authors declare no competing financial interests.

**Figure captions:**

**Figure 1 | Magnetic phases of MnP.** The *P-T* phase diagram includes ferromagnetism (FM), a double-helical order (H*a*-I) at low pressure [7], a new helical order (H*a*-II) discovered at high pressure in current work, superconductivity (SC), and paramagnetism (PM). Phase boundary data is adapted from Ref. [6] (open circles) with a reduction of pressure scale by a factor of 1.12 to match our x-ray measured H*a*-II phase boundary at 4 K (filled circle). (*P*, *T*) positions where the H*a*-II order was observed through magnetic scattering are marked (crosses). The presence of multiple ferromagnetic phases [25] is not distinguished here for clarity. (**inset**) Schematics of spin structures of three magnetic ground states, presented in a sequence of ascending pressure. The *n*-glide plane constraint between two helical orders in Ha-I is broken in the H*a*-II phase.

**Figure 2 | X-ray diffraction evidence of high-pressure helical order in MnP.** (**a-c**) Longitudinal (θ/2θ) line shapes of (2, 0, 0) lattice, and (1±*Q*’, 0,0) helical magnetic order. We set *a*>*b*>*c* for the lattice [7]. Lattice line shapes are instrument resolution limited for the whole pressure range, and can be fit to a Pseudo-Voigt form with a lattice coherence length exceeding 1500 Å. The magnetic peaks are significantly broadened, indicating a shorter correlation length of the helical spin order from ~310 Å at 3.2 GPa to ~70 Å at 6.4 GPa, about three times the pitch length of 24 Å. All peaks are fit with a Lorentzian form on a sloped background, which could be attributed to influence from spin fluctuations in the ordered phase [26]. However, our counting statistics are not sufficient to make a distinction from a Lorentzian-squared form, which results from disorder pinning [26]. The reduced background benefits from the use of a pair of wide-angle perforated diamond anvils [5]. Vertical dashed lines mark the commensurate (0.75, 0, 0) and (1.25, 0, 0) positions. Our instrument resolution is fine enough to indicate that the observed

magnetic pairs are mirror symmetric to the (1, 0, 0) order, but not commensurate. The presence of mirroring peaks around (100) indicates the *n*-glide plane constraint is broken for the spin arrangement at high pressure [7], although the (100) lattice order is still forbidden. (**d**) Above $P_c$=6.7 GPa, magnetic diffraction is no longer observed.

**Figure 3 | Scaled evolution of the magnetostriction and the magnetic phase boundary in MnP.** (**a-b**) Normalized lattice evolution under pressure, with $a(0)$ = 5.8959, $b(0)$ = 5.2361, and $c(0)$ = 3.1807 Å. $a(P)/a(0)$ and $b(P)/b(0)$ evolve slowly under pressure and are non-monotonic, while $c(P)/c(0)$ has a strong monotonic pressure dependence. The shapes of *a*, *b*, and $c(P)$ indicate large magnetostriction. Assuming that the lattice of a non-magnetic phase should evolve linearly over this pressure range, and that the low-pressure behavior can be modeled from extensions of the high-pressure lattice, the magnetostriction is then extracted by subtracting the estimated evolution of *a* and *c* (dashed lines in **a** and **b**). (**c**) Magnetostriction, expressed in both $\Delta c$ and $\Delta a$, can be scaled to magnetic phase transition temperatures $T_c$ and $T_N$ as a function of pressure. $\Delta c$ and $\Delta a$ are of different signs, indicating a positive Poisson ratio induced by the strength of the magnetic order.

**Figure 4 | Variable pitch length of the helical order as a tuning method for magnetically-mediated superconductivity.** (**a**) Superconducting transition temperature $T_c$ plotted as a function of helical wave vector $Q$ in selected 3*d* intermetallic compounds. Data for MnSi [16, 22], MnP [6], and CrAs [11-13] are collected from either the literature or current work. Superconductivity only was observed in pressure-induced disordered phases beyond the helical order and for $Q$ > 0.2 r.l.u., and is likely antiferromagnetically mediated. Ferromagnetically mediated superconductivity is expected to be at a lower temperature than its antiferromagnetic counterpart. [23, 24] (**b**) Schematic of a superconducting electron pair coupled through helical spin order in a projected planar view. The two sites of itinerant electron coupling are separated along the helical order by a half wavelength (λ/2), suggesting a singlet type of $d_{z^2}$-wave pairing. This scenario competes with superconductivity of a ferromagnetic type, while the nearly parallel local spin configuration always suppresses phonon-mediated superconductivity at a single-site [2]. For singlet superconductivity, λ of the helical order has to remain below the electron mean free path to allow their wave functions to overlap and preserve phase coherence.

**Extended Data Figure Captions**

**Extended Data Figure 1 | Reciprocal space scans in the low-pressure helical phase and high-pressure paramagnetic phase.** (**a**) Raw scans around (1, 1, 0) order at ambient pressure, showing both the lattice Bragg peak and a pair of non-resonant magnetic peaks associated with the helical spin order H*a*-I. Observed intensities of high pressure helical order is consistent with these magnetic intensities at ambient pressure. (**b**) At elevated pressure of 10.4 GPa above $P_c$, null longitudinal scans at the reciprocal space position of the H*a*-II order indicate the disappearance of the helix in the paramagnetic phase, while the lattice stays intact as exemplified by the profile of (2, 0, 0) order. Data in both panels were measured at $T$ = 4 K. Vertical error bars represent 1σ s.d. counting statistics. Solid lines are guides to the eye.

**Extended Data Figure 2 | Single crystal nature of the magnetic order at *P* = 5.28 GPa.** The single crystal nature of the magnetic order is proven by independent raw scans across the three dimensional reciprocal space for both (1-*Q*',0,0) and (1+*Q*',0,0) orders. The out-of-diffraction-plane transverse scan is dominated by the resolution function determined by the wide horizontal detector slits, while the in-plane transverse scan is intrinsic to the sample mosaic (FWHM ~ $0.1^o$) under pressure. Measurements were performed at $T$ = 4 K. Vertical error bars represent 1σ s.d. counting statistics. Solid lines are guides to the eye.

**Extended Data Figure 3 | Lattice constant ratio *a*/*c* as a function of pressure at *T* = 4 K.** The ratio evolves continuously across the critical pressure $P_c$, marked by the vertical dashed line, and is moving away from the value of 1.732 for the orthorhombic-hexagonal transition. The solid line is a guide to the eye.

**Extended Data Figure 4 | Magnetization *M*(*T*) and inverse susceptibility 1/*χ*'(*T*) at ambient pressure.** The measurement was performed in a SQUID based Magnetic Property Measurement System (Quantum Design) in a 100 Oe dc field,. $\chi'(T)$ was fit to the Currie-Weiss law above the ferromagnetic transition at 291K to extract a moment of 2.79 $\mu_B$/Mn. The measured Currie-Weiss moment is compared to the literature value of the saturated moment 1.3 $\mu_B$/Mn [7] in the high field and low temperature limit to provide a Rhodes-Wohlfarth ratio of 2.2.

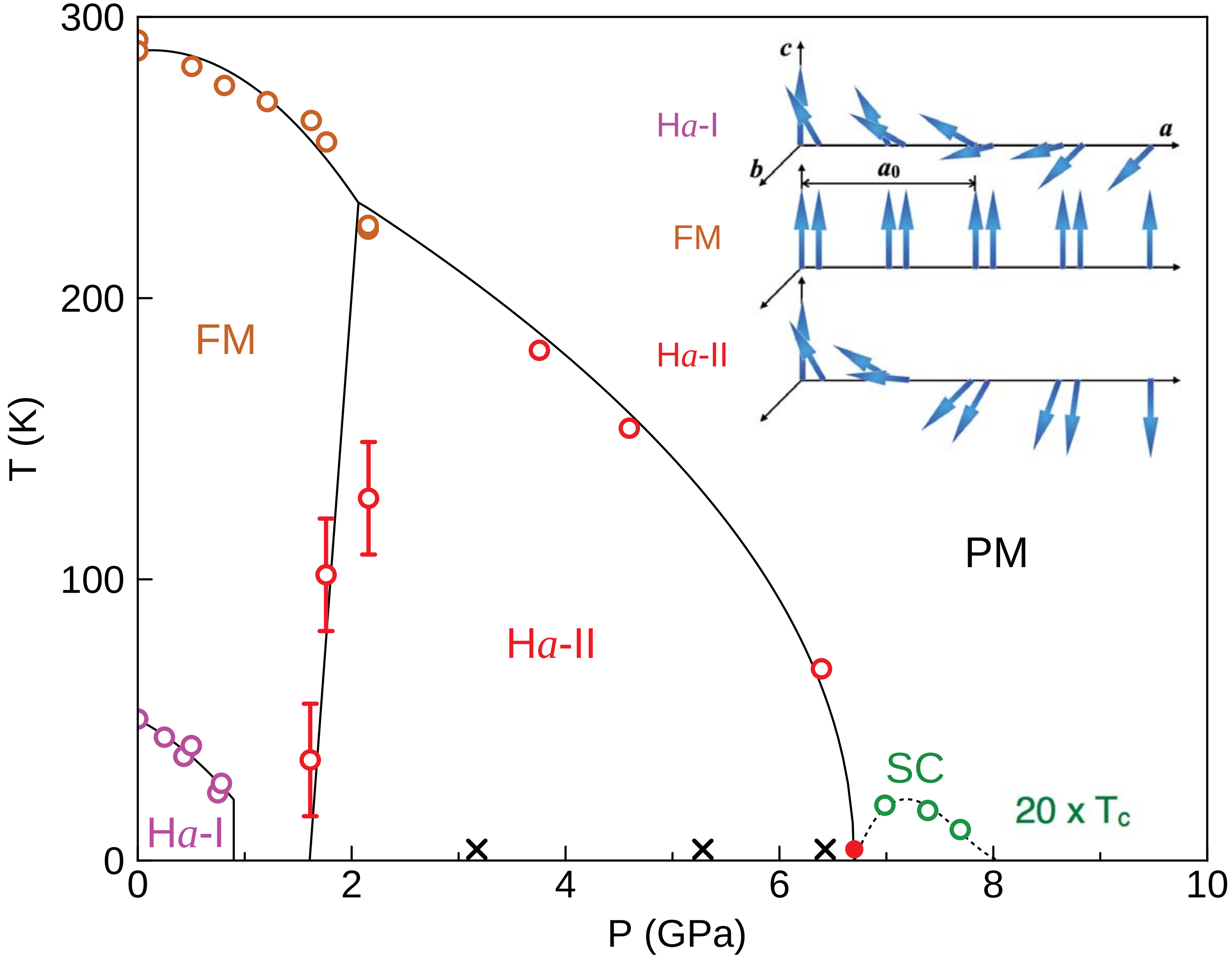

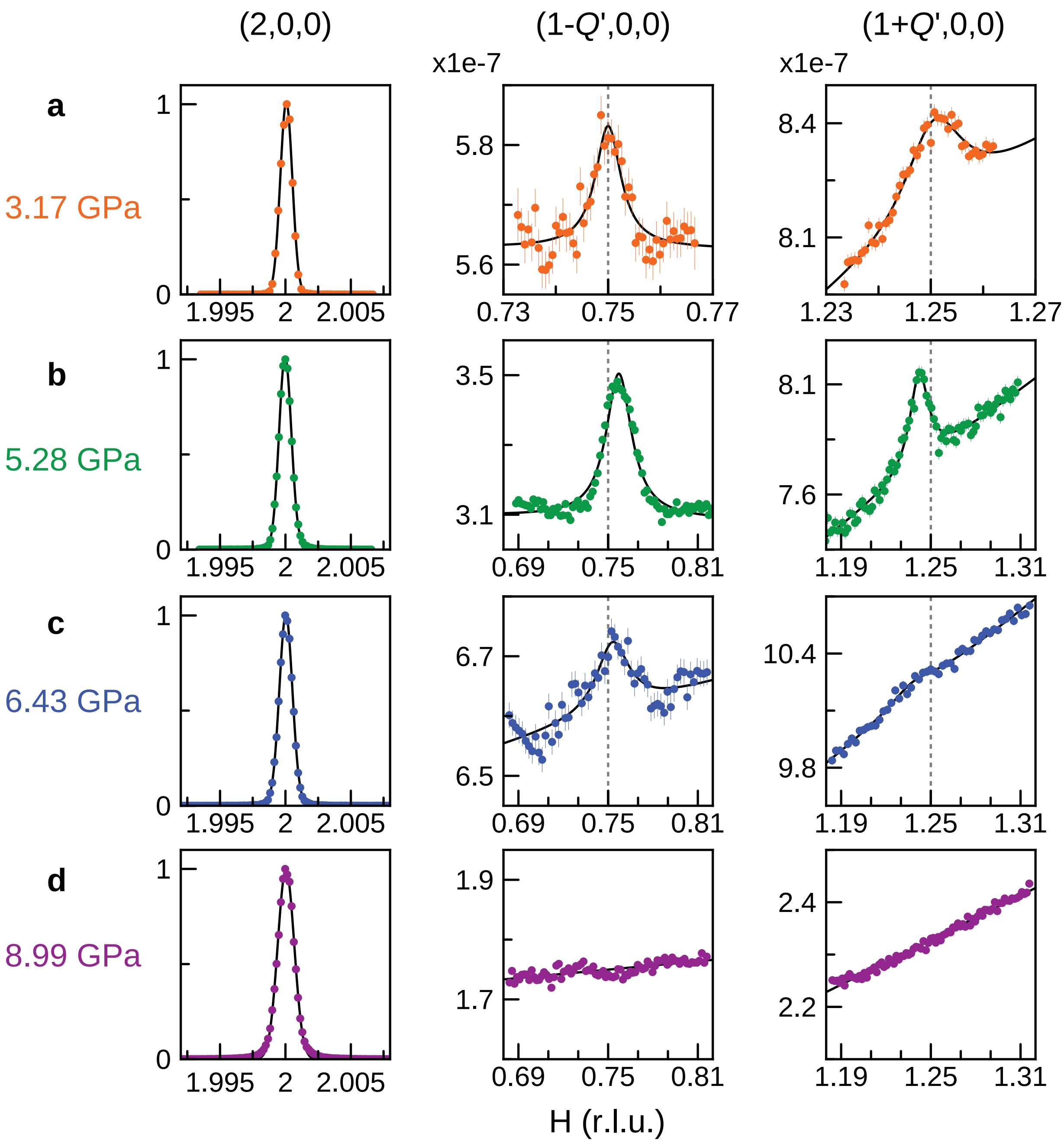

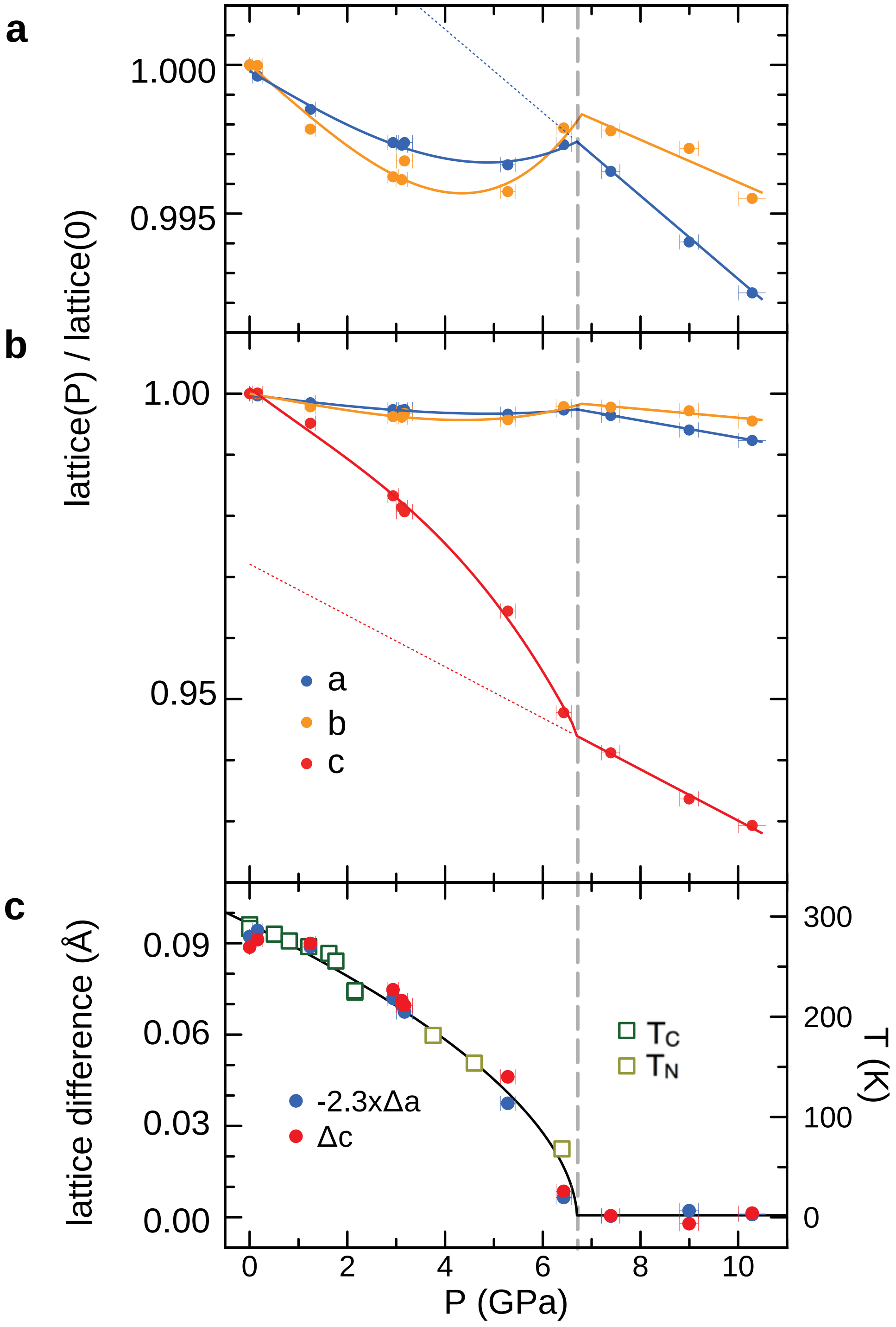

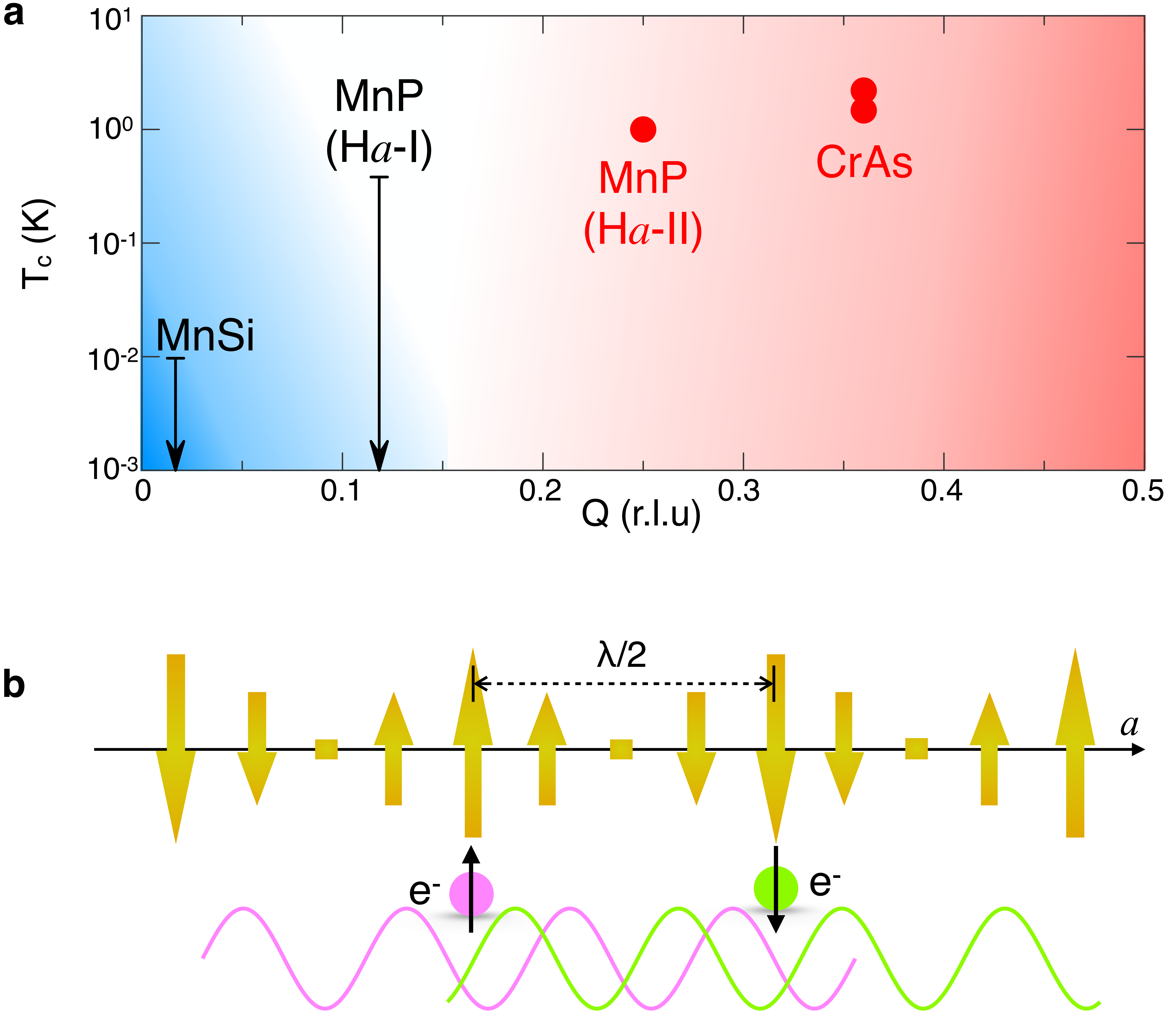

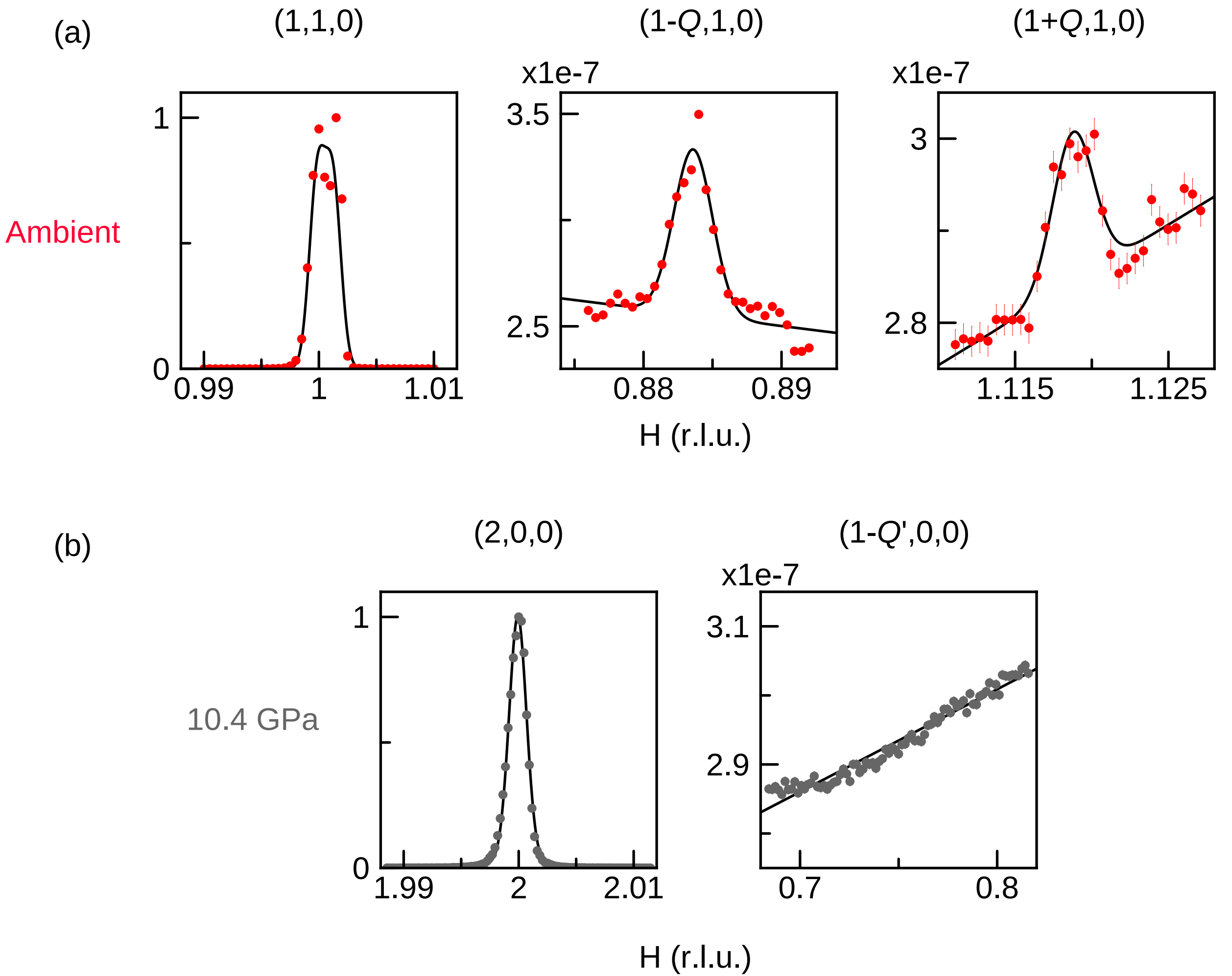

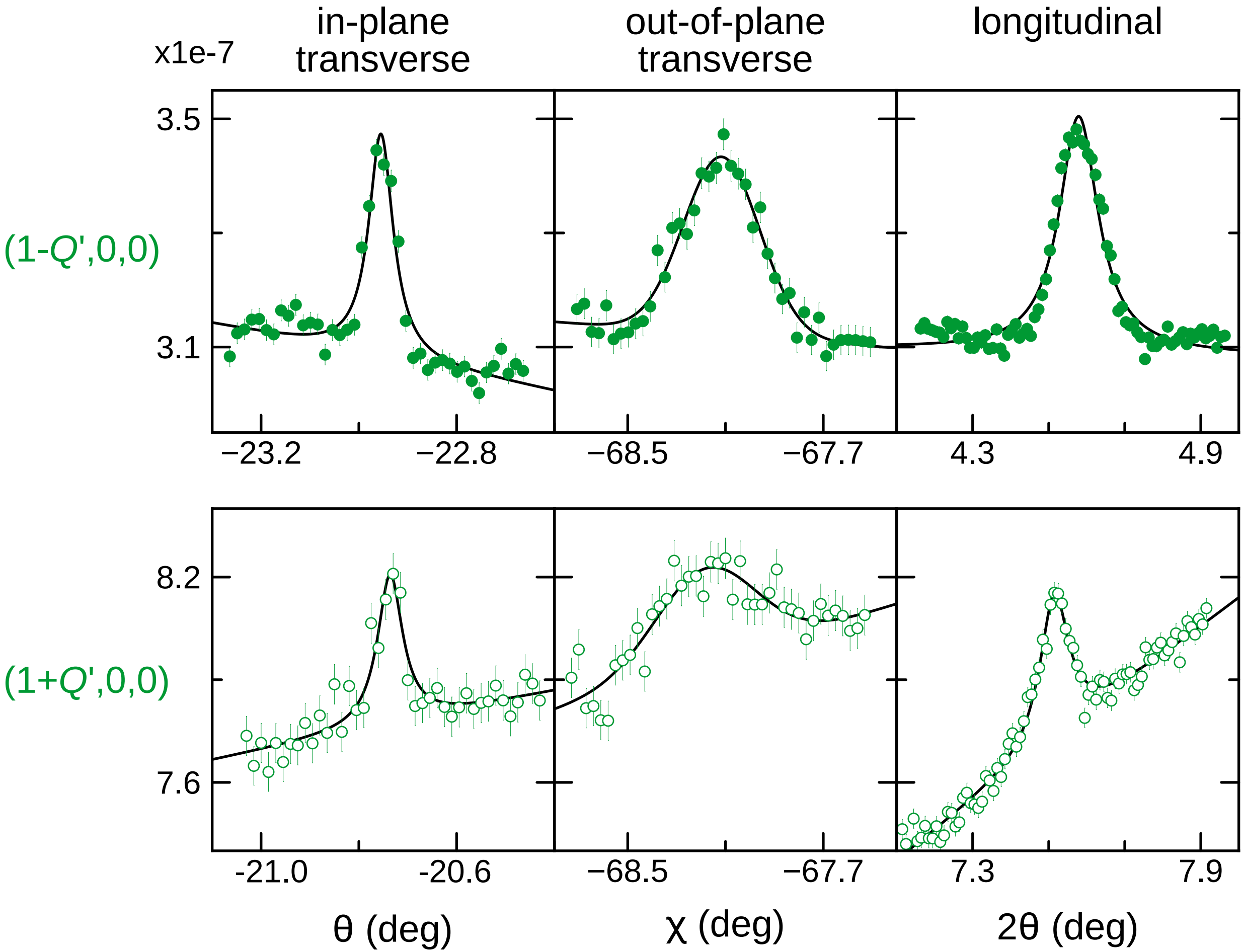

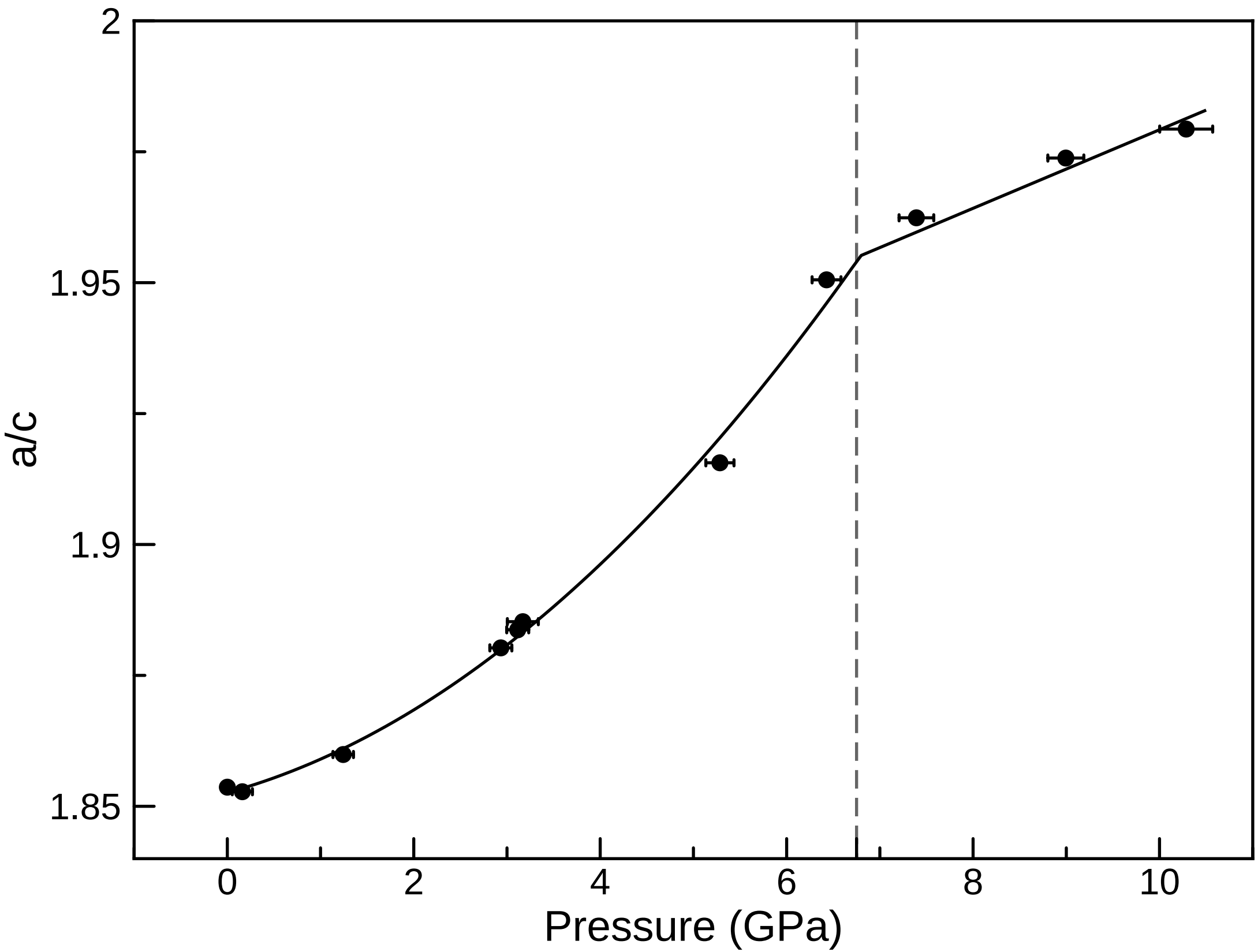

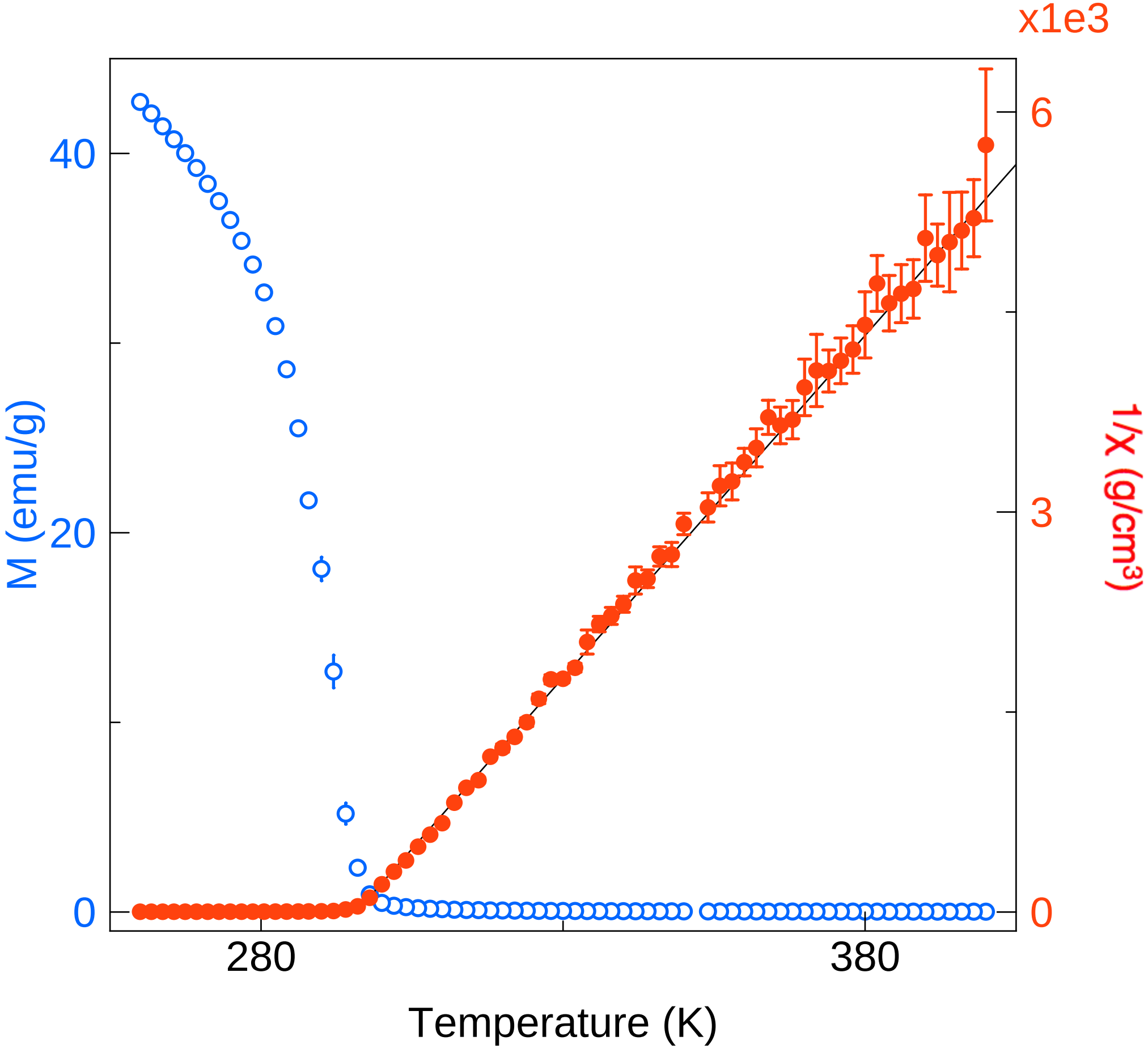